\documentstyle[11pt,paspconf,epsf]{article}

\begin{document}

\title{The Evolution of Quasars at High Redshift and Their Connection
with Galaxies}

\author{Patrick S. Osmer}

\affil{Department of Astronomy, The Ohio State University, 174 W. 18th
Ave., Columbus, OH 43210, USA, posmer@astronomy.ohio-state.edu}

\begin{abstract}

I review recent observations on the evolution of quasars, describe
new surveys for quasars at $z > 5$ and for quasars with $z > 3.3$ 
down to luminosities corresponding to L* galaxies, and note the
possible connection between the evolution of the star formation
rate in young galaxies and the evolution of quasars.

\end{abstract}

\section{Introduction}

I appreciate the opportunity to give the first talk at this meeting
and commend the Scientific Organizing Committee for bringing together
researchers on both quasars and galaxies.
I wish to mention some of the connections between the two fields that
may be valuable for present and future research.  Indeed, if quasars
are powered by supermassive black holes at the centers of galaxies,
the fields are not distinct.  Instead, they
are concerned with different stages of the formation of galactic-
sized objects in the universe.

In the case of quasars at $z > 3$, for example, we may be seeing the
results of the initial collapse of material at the center of a galaxy
and its fueling by the infall of additional matter.  The steep decline in
the space density of quasars from $z \approx 2$ to the present may
be caused by the decline in the fueling rate.  As galaxies complete
their formation, as their interaction rate presumably declines, and
as the universe continues to expand, the amount of matter falling
to the centers of galaxies likely declines, thus reducing quasar activity.
Alfonso Cavaliere will describe his work on this model later this morning.

In the meantime, the repair of HST, the advent of the Keck telescopes, and
the work on many different surveys have revolutionized observational
research on the formation and evolution of galaxies.  The abilities to
discover, to obtain images with resolutions near 0.1 arcsec, and to 
do spectroscopy of galaxies with redshifts greater than 3 and down
to 25th magnitude are truly changing our views of the universe.

At the same time, I think we quasar researchers are feeling rather 
distressed by all the excitement and developments with young galaxies.
Not only have galaxy researchers pushed into the realm formerly occupied
exclusively by quasars, but Franx et al. (1997) have reached beyond the
highest known quasar redshift to become the current recordholders in the
most-distant-known-object in the universe sweepstakes!  We will hear
from Marijn Franx about the lensed galaxy at $z = 4.92$ on Tuesday.  
Perhaps losing the redshift record will spur quasar researchers
to new depths, so to speak.  More seriously, the pursuit of both
galaxies and quasars at high redshift is important because it will 
tell us the epoch at which highly luminous objects first appeared
in the universe.

In this talk, I wish to discuss three topics:
\begin{enumerate}
\item Recent observational results on the evolution of quasars

\item Current surveys my collaborators and I are carrying out for
high-redshift quasars

\item The connection between the luminosity density of quasars and
the star formation rate in high-redshift galaxies
\end{enumerate}

As we shall see, the results of 1) provide motivation for 2), while 3) 
is of interest because of the similarity of the evolution of the quasar 
space density with redshift to the evolution of the star-formation
or metal-enrichment rate in young galaxies
(Madau et al. 1996, Connolly et al. 1997)

I will concentrate on optically selected quasars.  Subsequently 
Isobel Hook will describe the recent results for radio-selected quasars
and Stefano Cristiani will discuss his surveys for optical quasars.

\section{Recent Observational Results on the Evolution of Quasars}

When plotted on a linear scale against cosmic lookback time, $\tau$, the
space density of quasars shows a striking distribution with a peak near
$\tau \approx 0.85\,(q_0 = 0.5)$ and a width at half maximum of 
$\approx 0.1$ the age of the universe (Fig. 1).
\begin{figure}
\plottwo{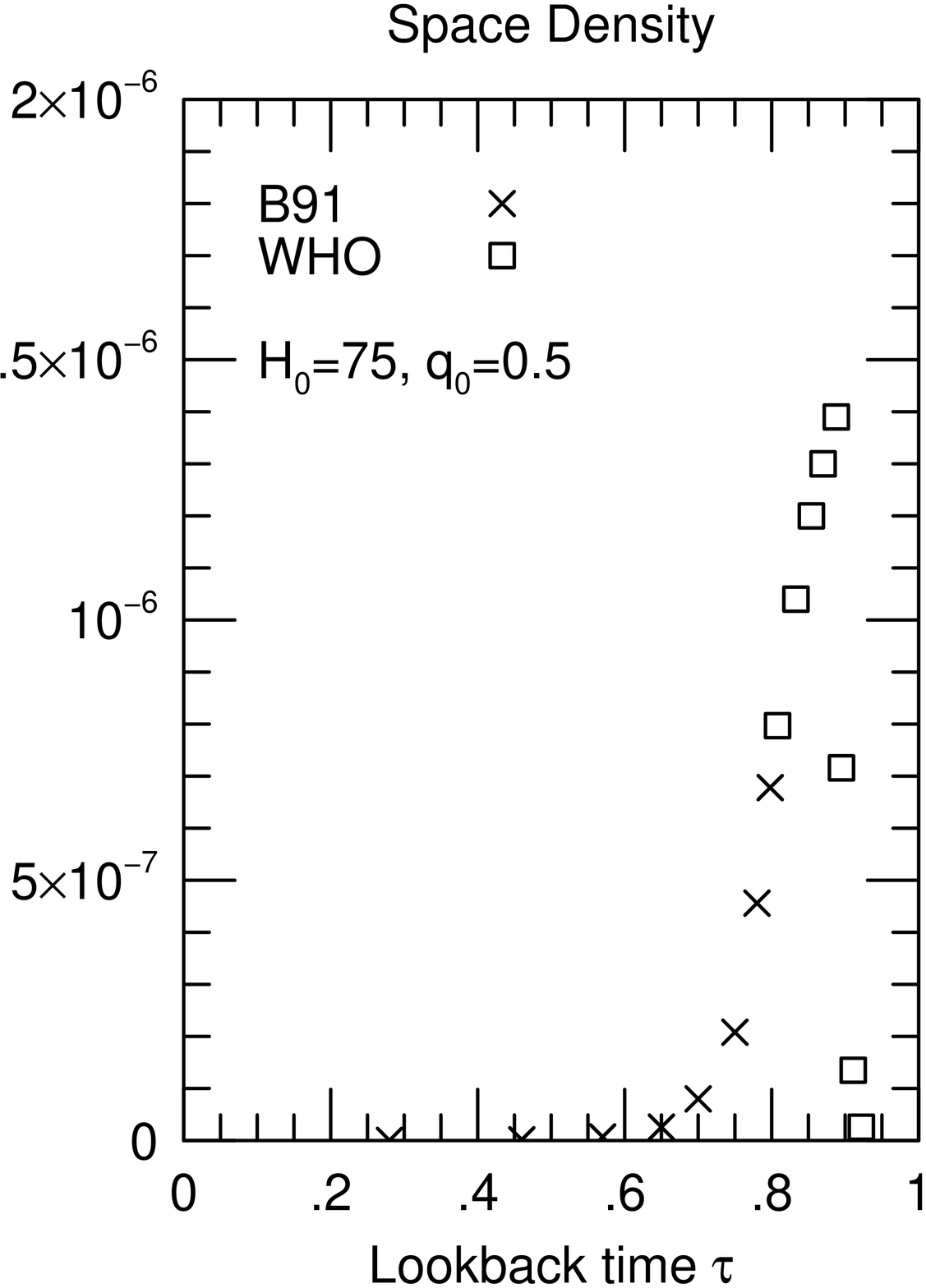}{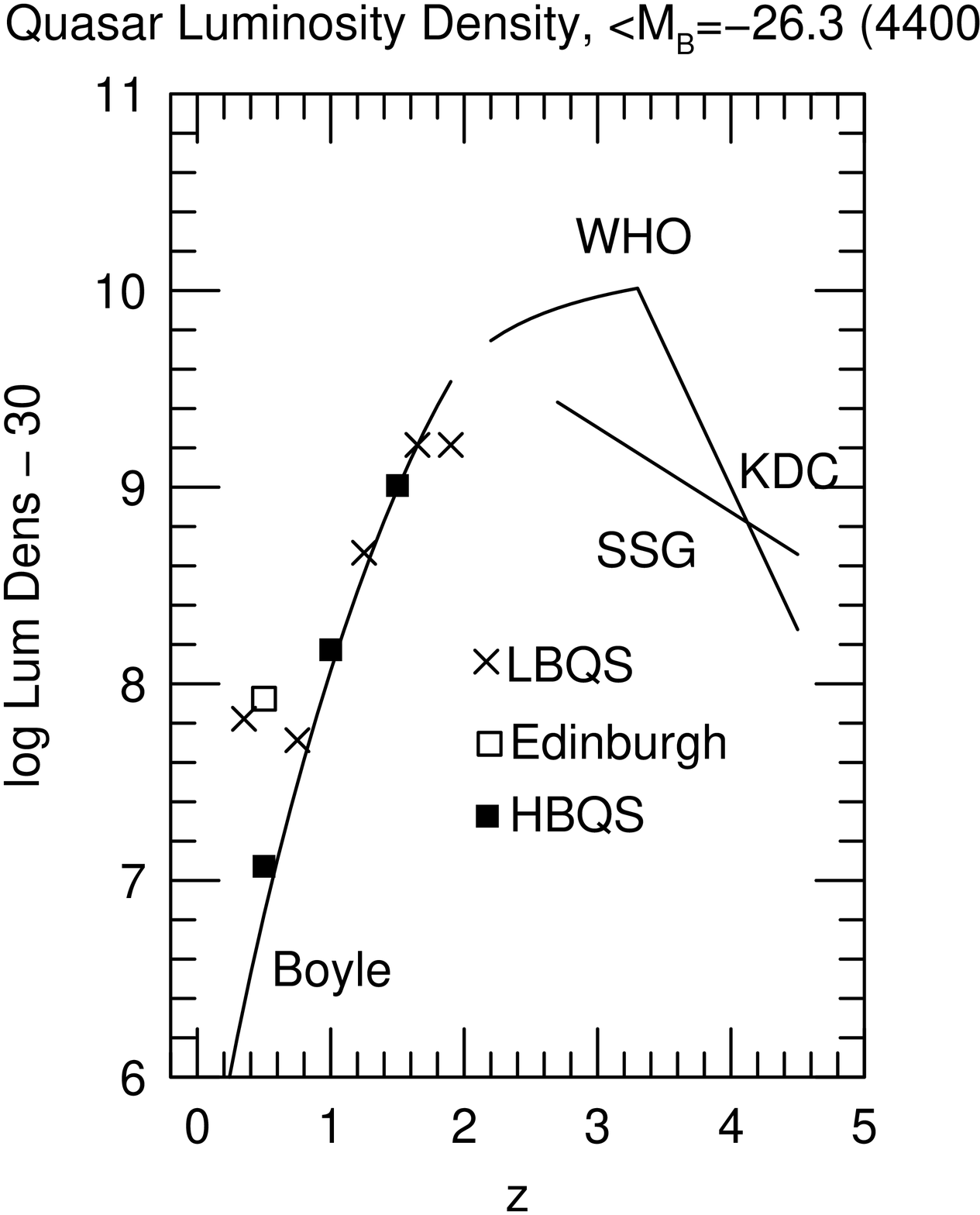}
\caption{(Left) Quasar space density, from Warren et al. (1994) and
Boyle (1991).  Figure 2.  (Right)
Quasar luminosity density for different surveys}
\end{figure}
\setcounter{figure}{2}
This plot continues to demand
questions such as: Was $\tau \approx 0.85$ the main epoch of quasar
formation? If so, why does the peak occur at that time and why does it
have such a narrow width?  Here I wish to concentrate on the observational
issues.

In Figure 2 I plot the logarithm of the luminosity density for various
quasar surveys against redshift.  I choose the luminosity density
in this case both for the comparison to be made in section 4 and
because this parameter characterizes how the integrated energy output
of quasars varies with redshift.

To illustrate the current observational state of quasar evolution,
Fig. 2 shows the parameterization from the AAT survey (Boyle, Shanks,
and Peterson 1988, Boyle 1991) for $0 < z < 2.2$ together with results
from the Large Bright Quasar Survey (LBQS, Hewett, Foltz, and Chaffee 1993),
the Edinburgh survey (Goldschmidt and Miller 1997), and the Homogeneous
Bright Quasar Survey (HBQS, La Franca and Cristiani 1997).

One of the major recent observational questions at low redshift and
bright magnitudes has been whether the Schmidt and Green (1983) results
and the Boyle parameterization under-represent the true population.  As we
see in Fig. 2, the LBQS and Edinburgh surveys do indicate a significant
correction is needed; the HBQS, less so.  My assessment is that a
correction on the order of a factor of several is now indicated at
$z \approx 0.4$ and thus the Boyle parameterization should be modified
for these redshifts and luminosities.  Note that this correction
is important, but it does not change the qualitative view of Fig. 1 because
the space density of quasars at $z \leq 0.5$ is still down by
two orders of magnitude from the peak.

At intermediate redshifts, $2 \leq z \leq 3$, we see from Fig. 2 that
the quasar luminosity densifty reaches its peak.  The calibrated surveys
of Warren, Hewett, and Osmer (1994, WHO) and Schmidt, Schneider, and
Gunn (1995, SSG) provide the main data in this interval. WHO find the
peak to be at $z \approx 3.3$, while SSG's results in combination with
Boyle's indicate that the peak is closer to $z \approx 2$.  I believe
additional work is needed to resolve this issue, which can be done with
surveys aimed particularly at the $2 < z < 3$ interval.  Again, the
uncertainty does not change the qualitative look of Fig. 1, but it does
indicate that the uncertainty in the location of the peak is $\approx 0.07$
times the age of the universe.

At higher redshifts, $z > 3$, we see in Fig. 2 that the WHO, SSG, and
KDC (Kennefick, Djorgovski, and de Carvalho 1995) surveys all show good
agreement at $z \approx 4.3$.  This provides, I think, persuasive evidence
that the apparent space density of optically selected quasars, which are
the bulk of the population, does decline for $z > 3$.  

However, Fig. 2
also shows that the WHO and SSG parameterizations are markedly different
in the slope of the decline, with WHO being much steeper.  This occurs for
two reasons, the different evolutionary decline with increasing redshift
{\em and} the different slopes of the luminosity function at the bright end.
When projected to $z > 5$, the WHO and SSG results give predictions for the 
number of quasars that differ by more than an order of magnitude.  Clearly,
additional surveys over the interval $3 < z < 5$ are needed to place
the quasar space density and the slope of its decline with redshift
on firmer ground.

\section{New Surveys for High-Redshift Quasars}

With the success of the multicolor technique for finding quasars
at high redshift (WHO, KDC, Irwin, McMahon, and Hazard 1993, IMH)
and the development of multi-CCD cameras, it is now possible to
carry out surveys that will either find quasars at $z > 5$ or set 
significant upper limits on their space density.  Because we do not
know well either what the slope of the luminosity function is at such redshifts
or the evolutionary decline with redshift,
it is important to survey a wide range of surface area and limiting 
magnitude.  Here I will describe the surveys being done by our groups;
other groups are also doing large programs, for example, Hewett and
Warren; Schmidt, Schneider, and Gunn; Irwin and McMahon, the Sloan
Digital Sky Survey.

To aid in visualizing the relation of the different surveys, Fig. 3
shows a plot of the log of their area {\em vs.} limiting magnitude.
\begin{figure}
\plotfiddle{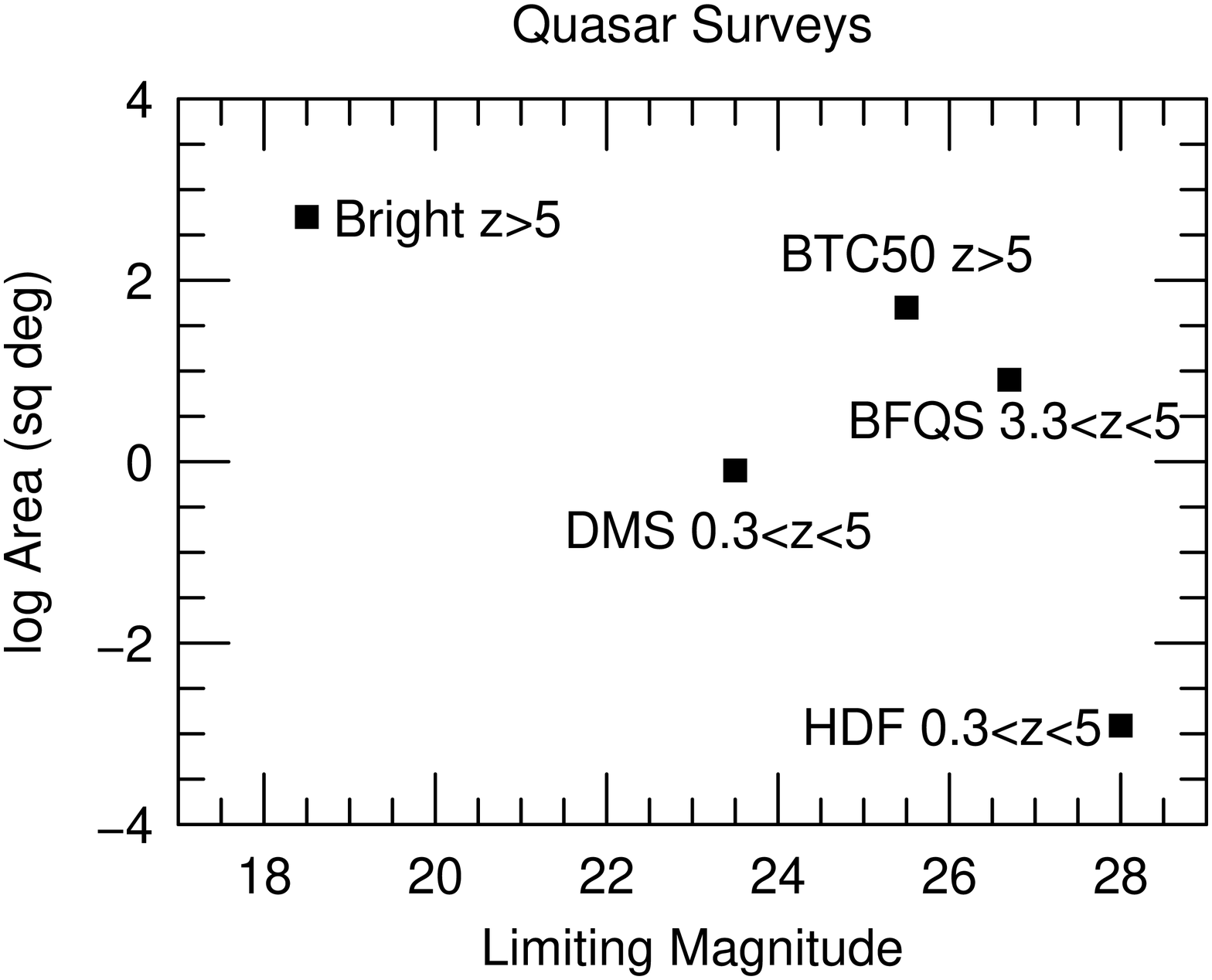}{2in}{0}{33}{33}{-144}{0}
\caption{Area and limiting magnitude of current surveys for quasars}
\end{figure}
The range in area is from 500 deg$^2$ with the Curtis and Burrell
Schmidts (in conjunction with the 2nd Palomar Sky Survey) to 
4.4 arcmin$^2$ in the Hubble Deep Field (HDF).  The range in limiting
magnitudes is roughly 18 to 28.  The main properties of the surveys
are described below:

\subsection{Deep Multicolor Survey (DMS)}

P. Hall, J. Kennefick, P. Osmer, R. Green, and S. Warren
are the investigators.  This survey covers 0.83 deg$^2$ in
U,B,V,R,I75,I86 to limiting magnitudes of 22.1 to 23.8 ($5\sigma$).
To date 55 quasars with $0.3 < z < 4.3$ have been confirmed 
spectroscopically, along with 43 compact narrow emission-line
galaxies (Hall et al. 1996a,b, Kennefick et al. 1997).  A catalog
of 21,375 stellar objects with positions, magnitudes in all six
bands, and error estimates for the magnitudes is being prepared
for publication.

\subsection{Bright $z > 5$ Survey}

This is being carried out by J. Kennefick, M. Smith, P. Osmer,
R. de Carvalho, A. Athey, and P. Martini.  It combines existing
data from the Digitized Palomar Sky Survey (DPOSS) with new data
taken at the Curtis and Burrell Schmidts in the $z (0.9\mu)$ band
to find $z > 5$ quasars to a limiting magnitude of $i \approx 18$.
Ly $\alpha$ emission in $z > 5$ quasars will make them very red
in $g - i$, and the $z$ band data help in eliminating late-type stars.
To date, data have been taken for 500 deg$^2$, and spectroscopic 
observations have been made of candidates selected from 300 deg$^2$.
No $z > 5$ quasars have yet been confirmed.

\subsection{BTC50 Survey}

E. Falco, P. Schechter, C. Kochanek, R. Green, M. Smith,
J. Kennefick, P. Osmer, P. Hall, and M. Postman
are the investigators for this survey, which is making use of the
BTC camera at CTIO to search for $z > 5$ quasars, gravitational
lenses, and distant galaxy clusters.  The goal is to cover 50 deg$^2$
in B,V,I, to limiting magnitudes of 25 to 26 in B and V and
24 in I.  Already 20 deg$^2$ of data are in hand.

\subsection{Big Faint Quasar Survey (BFQS)}

P. Hall, J. Kennefick, P. Osmer, R. Green, and M. Smith are also using
the BTC camera to cover 8 deg$^2$ in B,R,I to limits of 26.7, 25.7, and
24.6 to search for quasars with $3.3 < z < 5.0$ down to luminosities
equivalent to those of L* galaxies.  This survey will also be very useful
for studies of the evolution of field galaxies.

\subsection{Quasars in the Hubble Deep Field (HDF)}

A. Conti, J. Kennefick, P. Martini, P. Osmer, R. Pogge, and D. Weinberg
are searching for quasars and AGNs in the HDF by making use of the 
U,B,V,I images.  From numerical simulations, we find the
limiting magnitudes for classifying objects with the crude combine
images to be 25.2, 27.0, 26.8, and 25.8.  No clear $z > 4$ candidates
have been found so far.  One known emission-line galaxy with $z = 3.37$
has been independently recovered; it serves as a check on our modeling
procedures.  In addition, there are two possible UVX candidates.

\section{Quasars at $z > 5$?}

To date, I believe approximately $10^3$ deg$^2$ of sky have been surveyed
at brighter magnitudes ($I \approx 18$) by our group and by Hewett and
Warren, so far with negative results for quasars at $z > 5$.  While the
spectroscopy is not finished, we do have about 120 M-type stars, which
should yield a significant improvement in determining the lower end of
the stellar luminosity function.  In the meantime, Franx et al. (1997)
have discovered a lensed galaxy at $z = 4.92$.  Are we hitting the wall
on distant quasars?

In the spirit of Fig. 1 and 2, I show in Fig. 4 the currently known
\begin{figure}
\plottwo{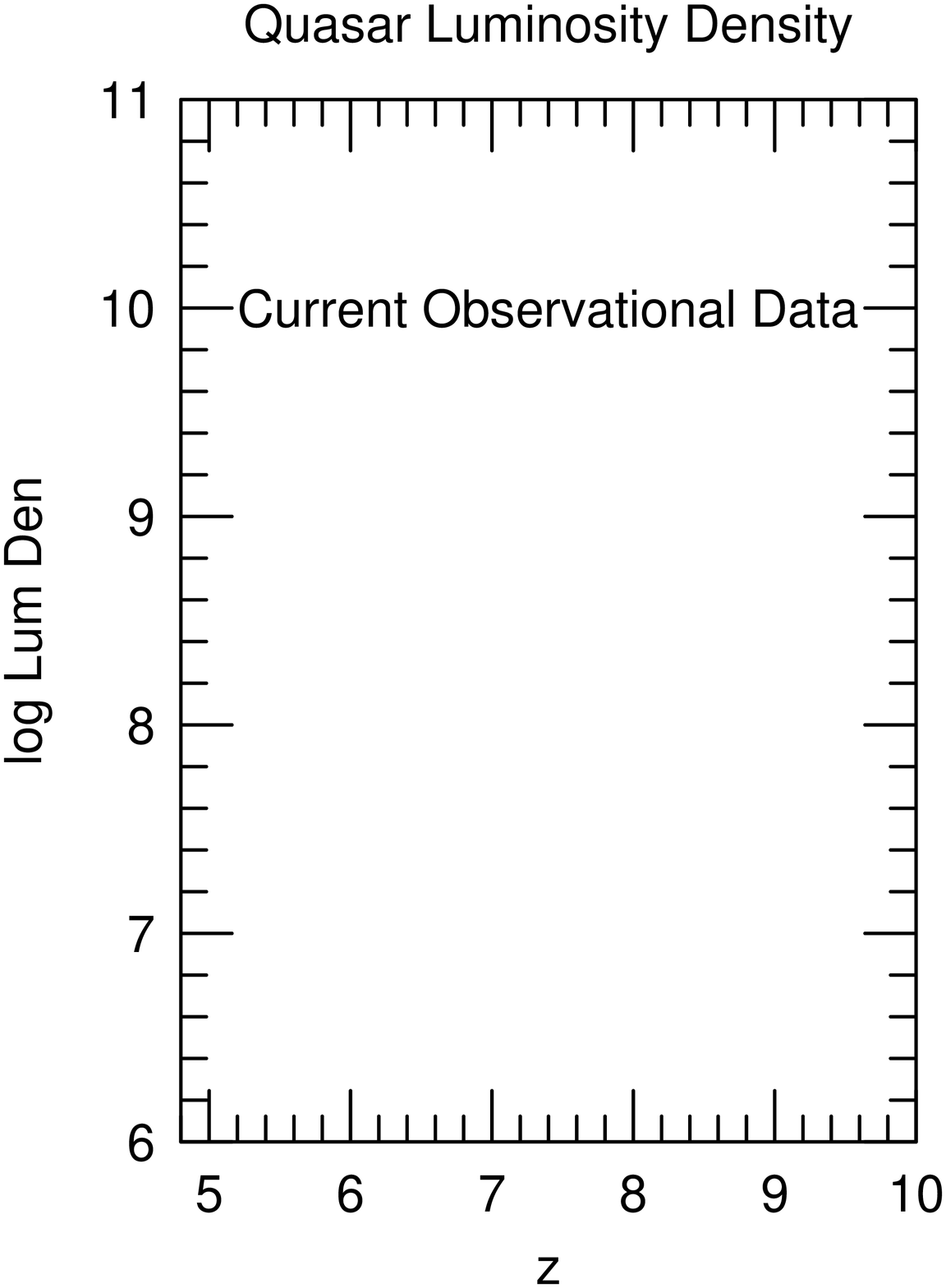}{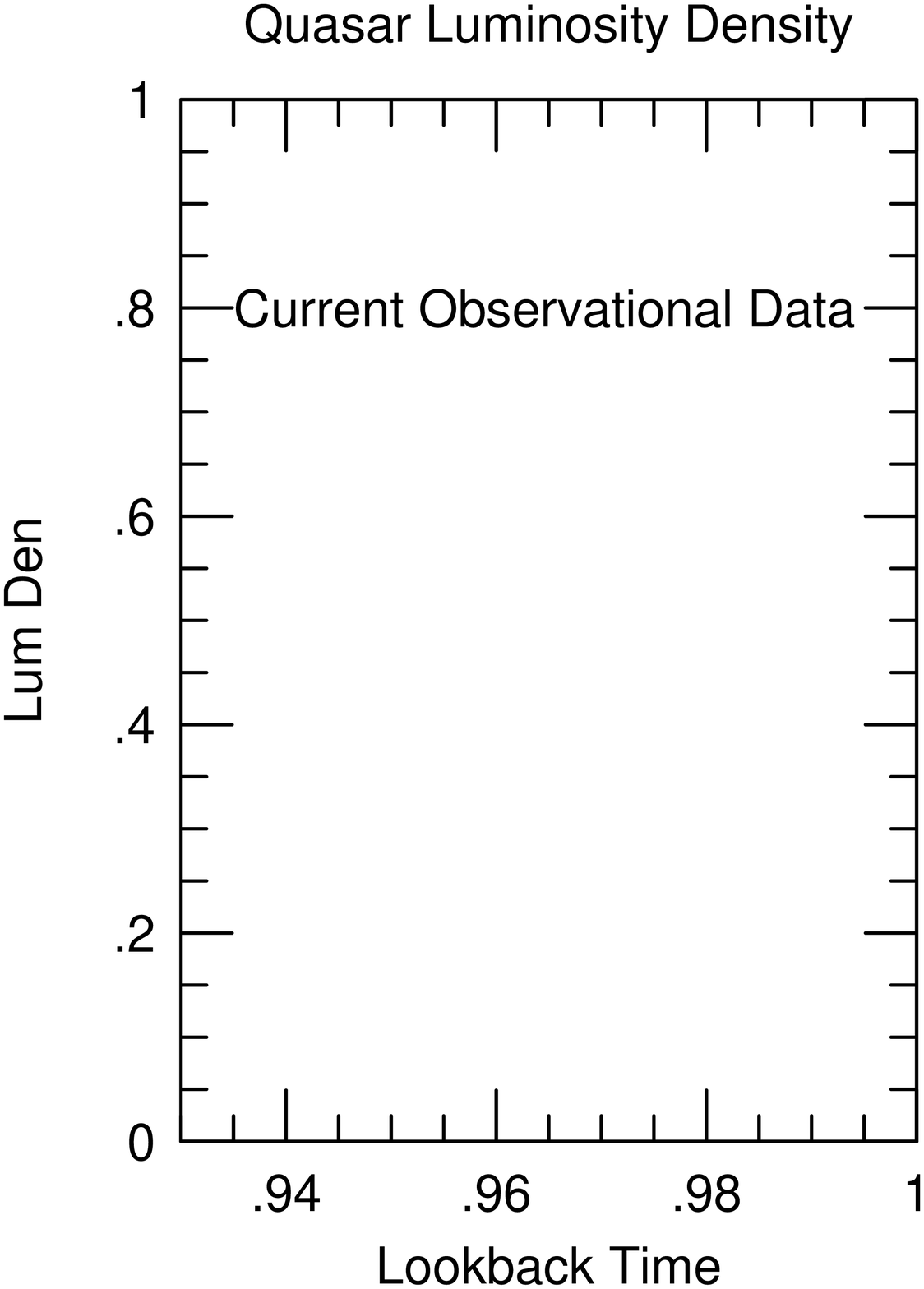}
\caption{Currently known space density of $z > 5$ quasars.  {\em Left}.  
Log of space density vs. redshift.  {\em Right}.  Linear plot of space
density vs. lookback time.}
\end{figure}
space density of $z > 5$ quasars, both in a log N vs. z and N vs. $\tau$
format.  Obviously, we need to fill in the diagram for the first 7\%
of cosmic time to understand when quasars first became visible.

\section{The Connection between Quasars and Galaxies}

The results of Madau et al. (1996, 1997) on the evolution of the
star-formation history of galaxies show evidence for the star-formation
rate (SFR) to increase with redshift by an order of magnitude from
$z = 0$ to $z = 1$ and then decline toward higher redshift.  The behavior
is reminiscent of but not identical to the evolution of the space
density of quasars.  The SFR peaks at a lower redshift and has a broader
distribution.  Is there a connection between the SFR in galaxies and
quasar evolution?

It is worthwhile to investigate the possible connection.  
The same processes responsible for the fueling of quasars may
well trigger star formation in the host galaxies.  Such processes
include galaxy interactions and instabilities.  Although the energy
sources for stars and quasars are different, the integrated ultraviolet
luminosity
density reflects the SFR for young stars in the case of high-redshift
galaxies and the fueling rate in the case of quasars.  
In addition, both luminosity
densities are directly observable quantities, and both have impacts
on the ionization state of the interstellar and intergalactic media
at high redshift.

To illustrate the connection, I use two forms
.
\begin{figure}
\plottwo{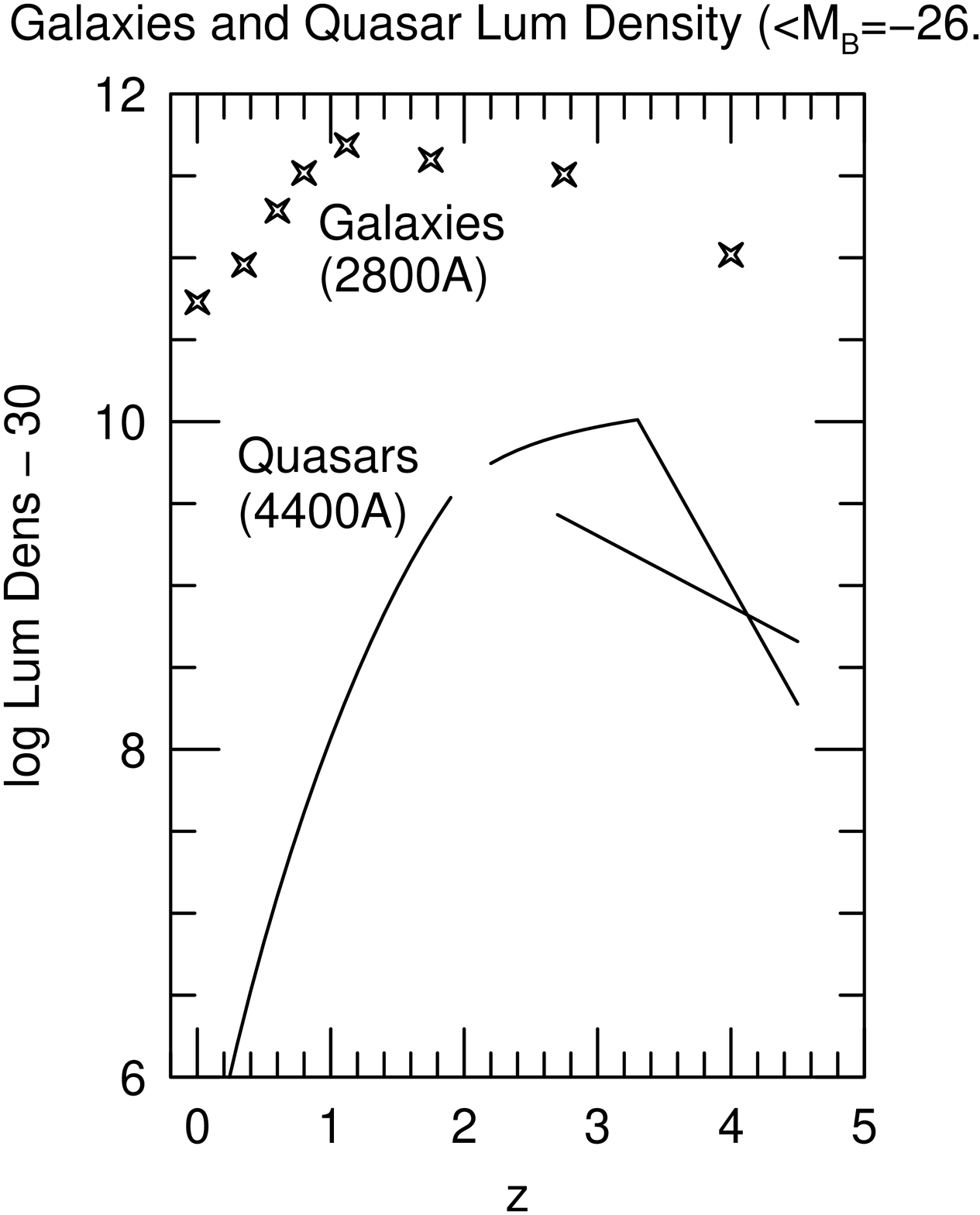}{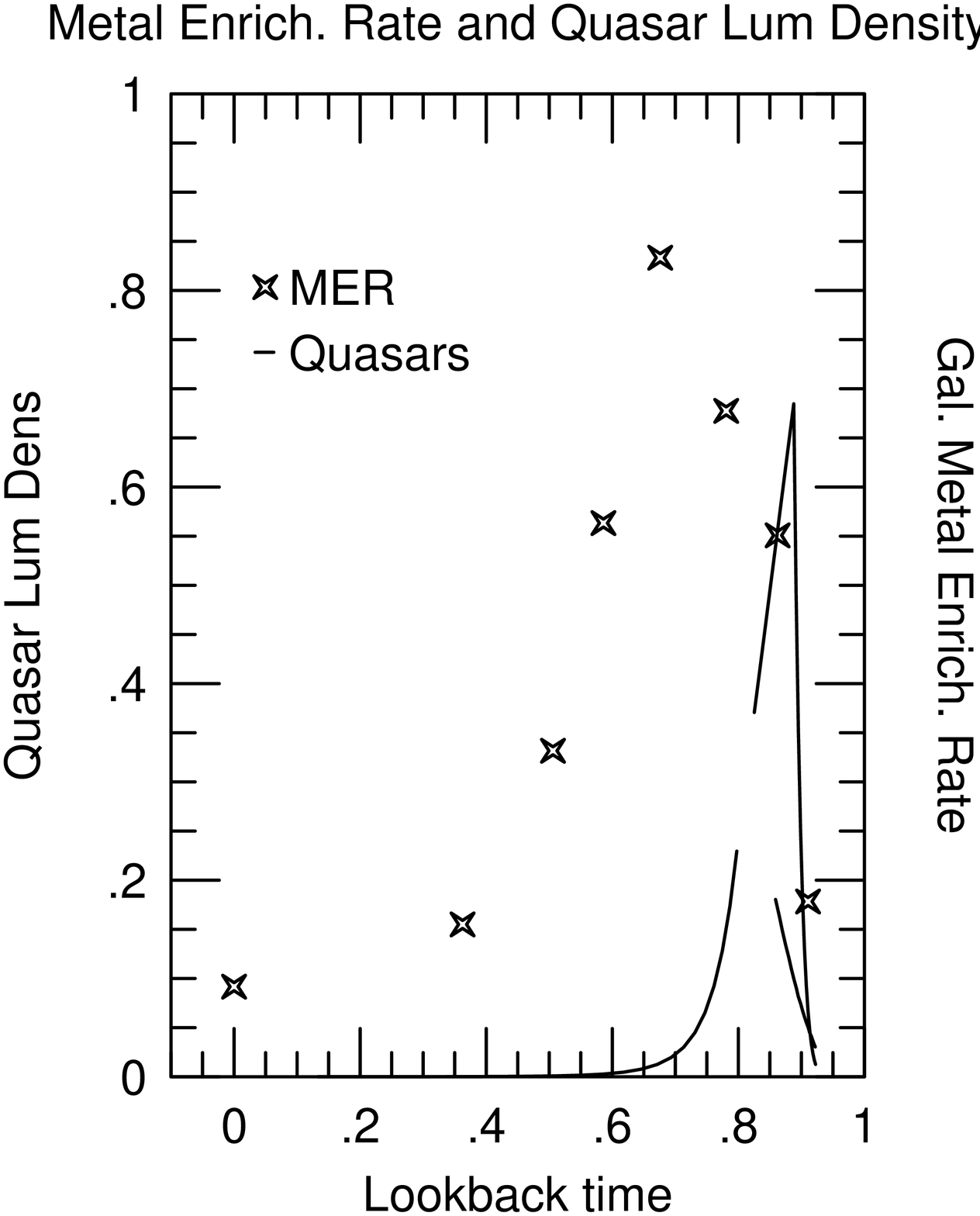}
\caption{{\em Left}.  A comparison of the luminosity density of
galaxies  ($2800\AA$) and luminous quasars ($4400\AA$). {\em Right}.
A linear plot of the metal enrichment rate and the quasar luminosity
density (arbitrary scale) vs lookback time.}
\end{figure}
The first, Fig. 5 (left),
is the log of the quasar luminosity density, adapted from Fig. 2,
and the log of the luminosity
density for the 2800$\AA$ emission of galaxies, adapted from
Connolly et al. (1997), {\em vs.} redshift.
We see that the luminosity densities
of the two classes of objects are within an order of magnitude of each
other.  When we consider that the stellar values have been integrated over a
Schechter luminosity function to represent the entire population of objects,
while the quasar values are only for luminous ($M_B < -26$) quasars,
the total quasar luminosity density may approach and even exceed the 
stellar luminosity density.  The second form, Fig. 5 (right) shows
a linear plot of the metal enrichment rate (MER) and quasar luminosity
density {\em vs}. lookback time.  The MER is shown because it is
less sensitive than the inferred
star formation rate to assumptions
about the slope of the initial mass function (Madau et al. 1996).  This form shows more directly
the offset in  cosmic time, 0.25 of the age of the universe, and the
greater breadth of the MER relative to quasars\footnote{Subsequent to
the meeting,
Boyle and Terlevich
(1997) independently noted this connection and discussed it in terms 
of the starburst model for quasars.}.

Although there are still uncertainties in the determination of the SFR and MER
for young galaxies and in the nature of quasar evolution,
Fig. 5 does provide a framework for observational mapping of the star 
formation history and the evolution of nuclear activity in young galaxies.
Continued observational work in turn should lead to improved 
understanding of how galaxies form and evolve.

\section{Summary and Future Work}

\begin{enumerate}

\item The main observational properties of the evolution of optically
selected, bright ($R \leq 20$) quasars are reasonably well established
for $0 < z < 4.5$.  However, we still need to determine better the
redshift at which the evolution peaks and especially the slope of its
decline toward higher redshift.

\item We now have the observational tools to observe both quasars and galaxies
to $z > 5$ and to magnitudes $\geq 25$. Thus, we have the capability
to map observationally their evolution down to L* luminosities at
$z \approx 5$ and 0.1 L* at $z = 2$.  The data will be important
to improving our theoretical understanding of galaxy formation and
evolution.

\item There is reason to expect the discovery of galaxies and quasars
at $z > 5$ (the increment in cosmic time is only 0.001 of the age of the
universe compared to the most distant object now known ($z = 4.92$).  
\end{enumerate}

Do quasars exist at $z > 5$?  Perhaps we will have an answer by the
time of the next meeting on this topic.  In any case, exploring the
properties and evolution of both galaxies and quasars at high redshift
will be one of our grandest adventures.

\acknowledgements This work was supported by NSF grant AST-9529324.


\end{document}